\documentclass[%
 preprint, 
 amsmath,amssymb,
 aps, physrev,
]{revtex4-2}

\usepackage{textcomp}
\usepackage{graphicx}% Include figure files
\usepackage{dcolumn}% Align table columns on decimal point
\usepackage{eurosym}
\usepackage{bm}% bold math
\usepackage{braket}
\begin{document}

\preprint{APS/123-QED}

\title{\textbf{A portable LED-based diamond magnetometer for outreach and teaching labs} 
}% 

% \affiliation{%
%Graduate School of Advanced Science and Engineering, Hiroshima University, Kagamiyama 1-3-1, Higashi Hiroshima 739-8530, Japan
%}%

\author{Hollis Williams}%
 \email{Current address: Department of Mathematics and Statistics, University of Exeter, Exeter EX4 4QF, UK}
  \altaffiliation[]{holliswilliams@hotmail.co.uk}
 \author{Alex Newman}%
\author{Stuart Graham}%
\author{Colin Stephen}%
\author{Gavin Morley}%
\email{gavin.morley@warwick.ac.uk}
\affiliation{%
 Department of Physics, University of Warwick, Coventry CV4 7AL, United Kingdom
}%

% \author{Hollis Williams}
% \altaffiliation[]{Hollis.Williams.1@warwick.ac.uk}
 
%%\author{Alex Newman}%
%\author{Stuart Graham}%
%\author{Colin Stephen}%
%\author{Gavin Morley}%
% \email{Contact author: Second.Author@institution.edu}
%\affiliation{%
% Department of Physics, University of Warwick, Coventry CV4 7AL, United Kingdom
%}%

%\affiliation{
% First affiliation for this author
%}%
%\affiliation{
% second institution for this author
%\affiliation{%



\begin{abstract}

We present a compact, low-cost version of an NV center diamond magnetometer which replaces the standard green laser with a high-power LED. This modification improves safety, reduces cost, and allows the green excitation and red photoluminescence to be viewed directly during demonstrations. The device is simple to assemble and suitable for outreach activities and undergraduate laboratories. We show that it can produce ODMR spectra and respond to nearby magnetic objects, with a sensitivity on the order of 
1 $\mu$T/$\sqrt{\text{Hz}}$.  Supplementary material provides details of the construction and suggestions for student investigations to support use in teaching laboratories.

\end{abstract}

\maketitle

%\tableofcontents

\section{\label{sec:level1} Introduction}

The rapid growth of quantum technology has created a demand for quantum teaching labs and demonstrations at undergraduate level.  NV (nitrogen vacancy) centers in diamond are appealing for such demonstrations, because they are stable qubits at room temperature and can be used to study a range of quantum phenomena \cite{sewani, yuan}.  They also have various applications in quantum computing and magnetic sensing.  Through optically detected magnetic resonance (ODMR), students can visualize spin-state dynamics, Zeeman splitting, and basic magnetometry principles. 

Several recent works have outlined teaching experiments based on the physics of the NV center and provide simplified designs suitable for undergraduate-level laboratories \cite{zhang}.  Stegemann et al. demonstrated a modular, mostly 3D-printed ODMR apparatus which uses micro-sized NV diamonds and inexpensive electronics, enabling a complete experiment for well under \euro{}250 \cite{steg}.  Similarly, the open-source Uncut Gem platform from Quantum Village provides a modular NV magnetometer which relies on easily available components and minimal optics \cite{village, village2}.  These designs offer highly affordable routes for observing ODMR spectra and so represent the cheapest possible options for introducing the physics of the NV center into teaching laboratories.

In this Note, we present a portable NV magnetometer which uses a high-power green LED instead of a laser, allowing for safe illumination of the diamond with optical powers of order 370 mW.  This magnetometer prioritizes safety, portability, reliable performance, and direct visualization for outreach and undergraduate settings.  The high excitation power produces fluorescence which is readily visible to the naked eye, making the system particularly effective for public demonstrations and instructional settings.  Although it does not reach the affordability of the very low cost modular options which are available in the literature, the magnetometer which we present here is tailored for teaching and outreach environments where clarity of the fluorescence signal and ease and reliability of the demonstration are essential.  

 Low power (few-milliwatt) 532 nm laser modules can also be used in teaching lab implementations of ODMR experiments.  However, LED illumination requires no laser safety procedures or alignment, making it particularly suitable for outreach demonstrations and environments with minimal supervision.  Despite its simplicity, the LED device produces clear ODMR spectra and can detect changes in the local magnetic environment with a sensitivity of around 1 $\mu$T/$\sqrt{\text{Hz}}$.  We provide a concise description of the operating principles and demonstration capabilities of the apparatus, with detailed assembly instructions, schematics, and example data provided in the Supplementary Material.

\section{\label{sec:level1}  Concept and Demonstration}

The central idea of the device is to use a high-power green LED (approximately 370 mW of optical power) to excite a millimetre-scale diamond containing a high concentration of NV$^-$ centers, and to detect the resulting red photoluminescence (PL) while applying microwave radiation near the NV spin resonances. A simplified schematic of the optical and electronic layout is shown in Fig. 1 (left). The LED output is fed into a short hexagonal light-mixing rod, which homogenizes the illumination and delivers it to a right-angle prism. The prism directs the green light onto the diamond mounted on the printed circuit board, whilst simultaneously collecting the red photoluminescence and returning it through the rod to a long-pass filter and amplified photodiode. Since the green excitation and red fluorescence are both easily visible to the naked eye, students can directly observe the physical processes responsible for the ODMR signal.

To clarify the optical and mechanical interface on the circuit board, Fig. 1 (upper right) provides a detailed view of the prism-rod-diamond assembly. The diamond is bonded to a reflective metal pad on the board, the prism is placed on top of the diamond, and the face of the mixing rod is aligned so that it presses against the face of the prism. This compact geometry provides efficient light collection without the need for lenses and keeps alignment requirements minimal.  The board incorporates a printed single-turn loop antenna which provides the oscillating magnetic field required for ODMR. The loop is etched into the top copper layer and is driven from the SMA connector through a short copper transmission line, as shown in Fig. 1 (lower right). The diamond is positioned directly above the loop to ensure strong near-field coupling. An excerpt of the top copper layer is included in the figure for clarity, and the full Gerber files are available in the Supplementary Material.

\begin{figure}[b]
  \centering

  \begin{minipage}[t]{0.4\linewidth}
    \vspace{0pt}
    \includegraphics[width=\linewidth]{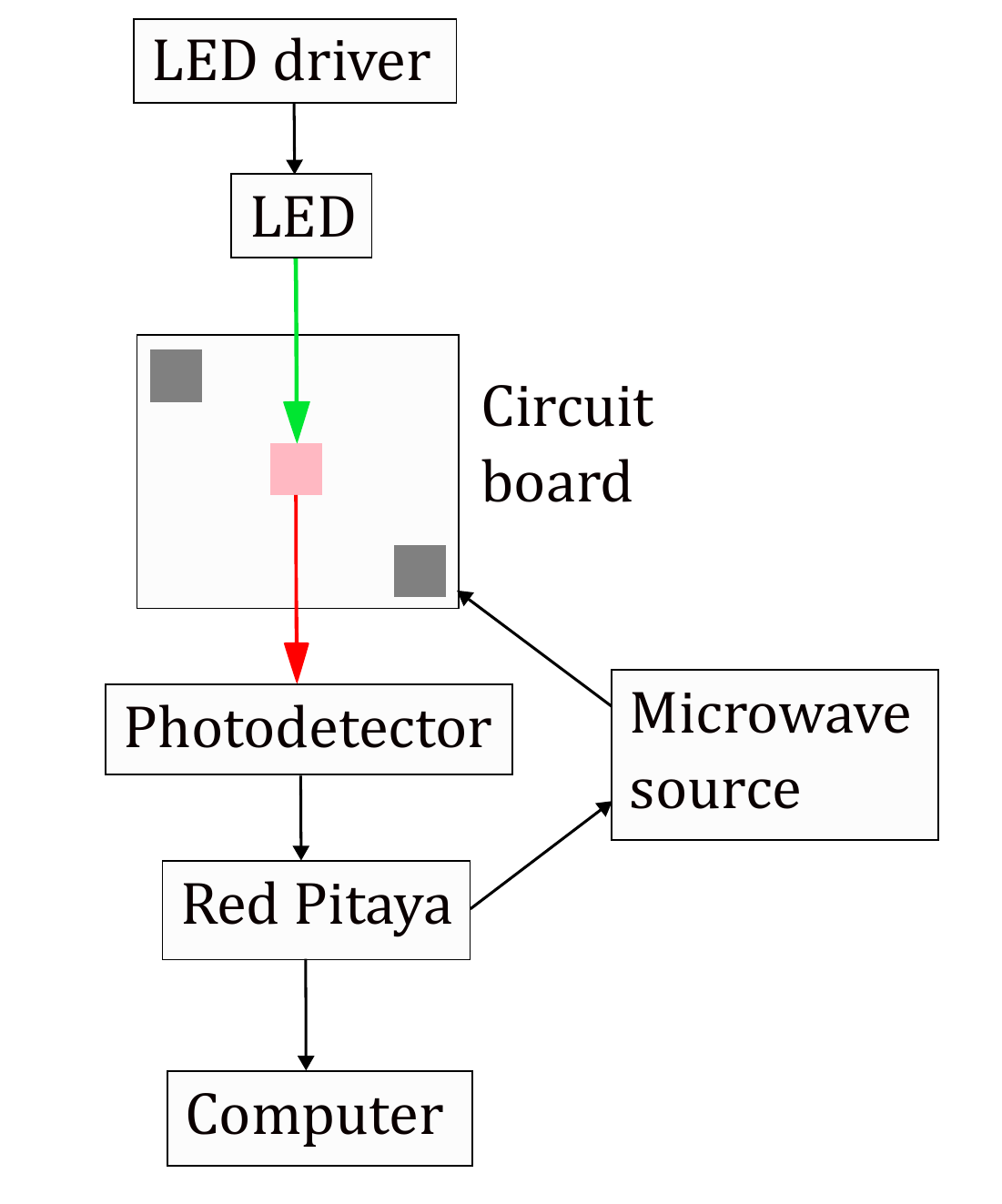}
  \end{minipage}%
  \begin{minipage}[t]{0.4\linewidth}
    \vspace{0pt}
    \includegraphics[width=\linewidth]{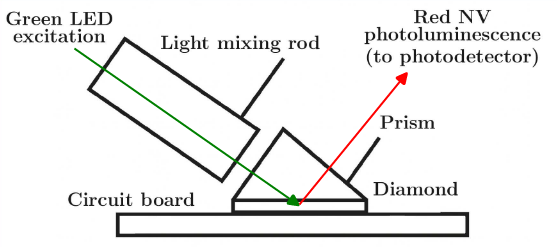}\\
    \includegraphics[width=\linewidth]{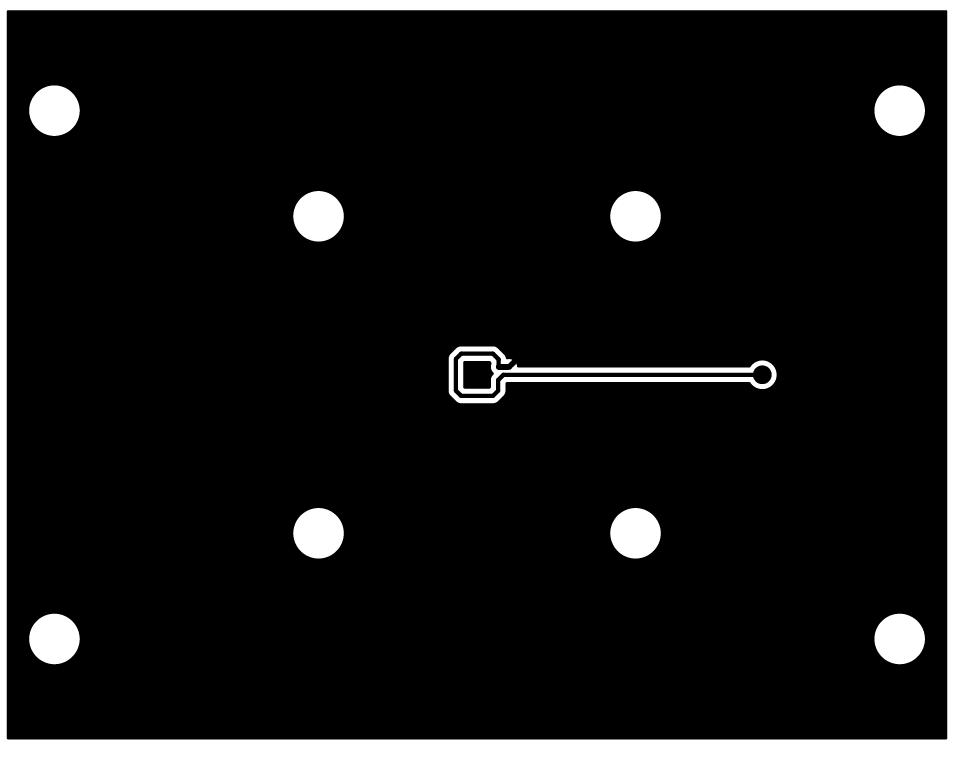}
  \end{minipage}

\caption{(Left) Schematic of the experiment.  The gray squares denote the two block magnets attached to the circuit board and the pink square denotes the NV diamond.  (Top right) Schematic showing the optical layout (not to scale). The green arrow indicates the excitation path from the LED to the diamond. The red arrow indicates NV photoluminescence emitted by the diamond and directed toward the photodetector (Bottom right) Excerpt of the top copper Gerber layer of the printed circuit board showing the integrated microwave loop used for ODMR. The circular pad on the right connects to the SMA input. A straight copper trace routes the microwave signal to the printed single-turn loop in the middle of the board, which generates the oscillating magnetic field to drive the NV spin transitions. The diamond sits directly above this loop during operation. Full PCB design files are provided in the Supplementary Material.   }

  \label{fig:led}
\end{figure}

The Red Pitaya–based digital lock-in amplifier used in this work is implemented with open source software running on a host computer. In the present experiment, students interact with this system through a graphical user interface (GUI) that provides real-time graphical visualization of the demodulated ODMR signal, without requiring prior knowledge of digital signal processing or lock-in implementation.  The interface can be operated in several ways.  For ODMR spectroscopy, the microwave frequency can be swept whilst the magnitude of the lock-in response is recorded as a function of frequency, allowing ODMR resonances to be observed. Alternatively, the microwave frequency can be fixed at a selected working point and the in-phase (X) and quadrature (Y) lock-in outputs monitored as a function of time, enabling changes in the magnetic environment to be detected.

In this work, the GUI is used primarily as a qualitative visualization tool, allowing students to explore the ODMR response and its dependence on experimental parameters such as microwave frequency range, modulation depth and frequency, lock-in time constant, and LED drive current (where available). Changes to these parameters are reflected in the displayed signal, allowing students to investigate their influence on ODMR contrast, linewidth, signal-to-noise ratio, and magnetic field response. The simplified optical design eliminates the need for laser alignment, allowing students to focus on signal interpretation, data analysis, and the physics of magnetic resonance rather than optical troubleshooting.

The magnetometer has been used successfully in outreach activities. The entire setup packs into two small plastic boxes and can be reassembled in a few minutes, shown in Fig. 2 (left).  Fig. 2 (right) shows that both the green LED excitation and red photoluminescence can be seen with the naked eye.  At a recent event in our Physics Department, students viewed the PL signal on a laptop whilst magnetic objects were brought near the sensor. For simple demonstrations, it is sufficient to observe the raw PL signal, which changes visibly in response to nearby magnetic objects.  Fig. 3 shows the demodulated PL signal when a steel Allen key was moved close to the magnetometer.  It can be seen that a spike forms as the object is brought near, giving a compelling demonstration that the device functions as a magnetometer.

\begin{figure}[b]
  \centering

  \begin{minipage}[t]{0.70\linewidth}
    \vspace{0pt}
    \includegraphics[width=\linewidth]{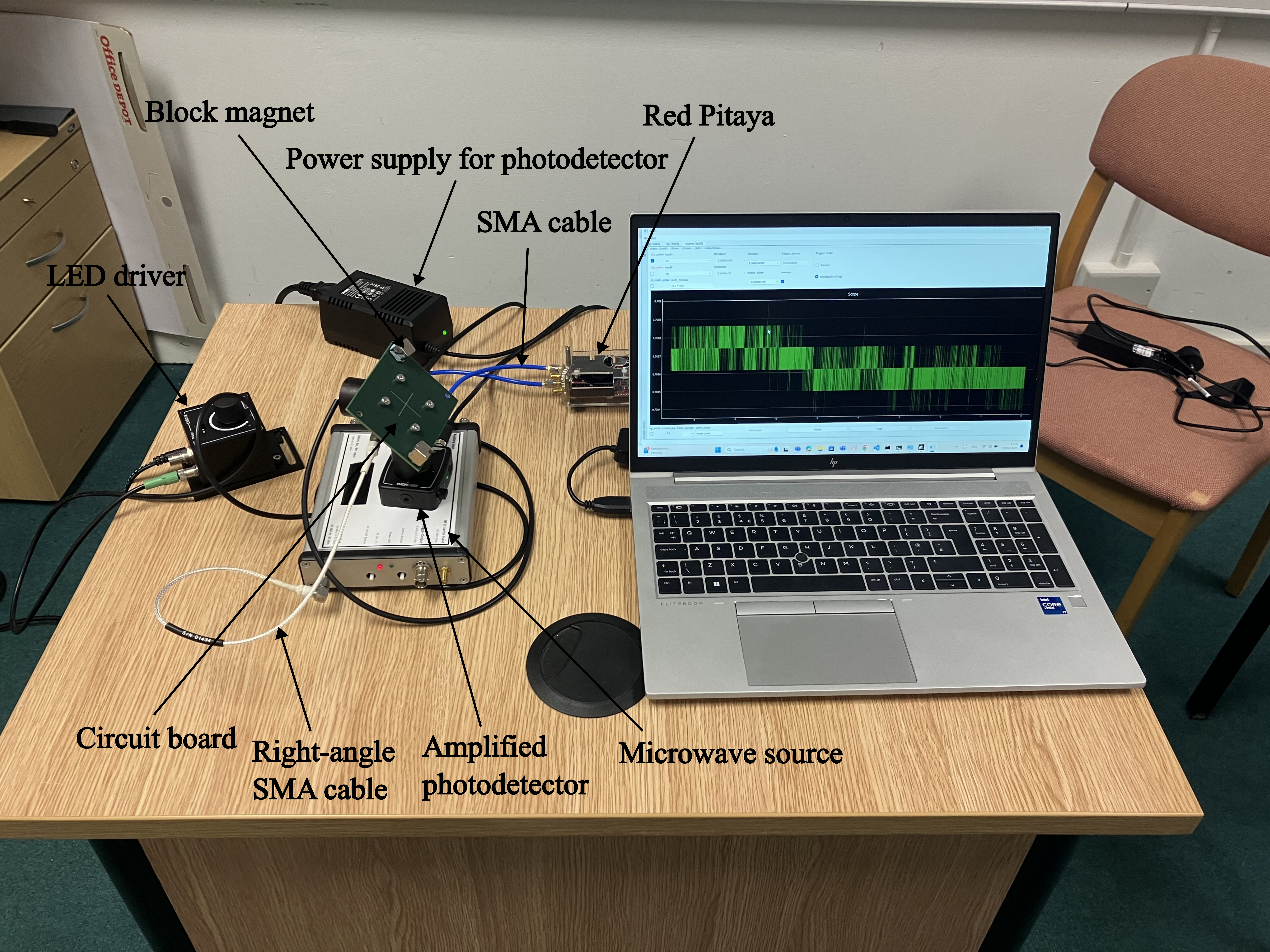}
  \end{minipage}%
  \begin{minipage}[t]{0.257\linewidth}
    \vspace{0pt}
    \includegraphics[width=\linewidth]{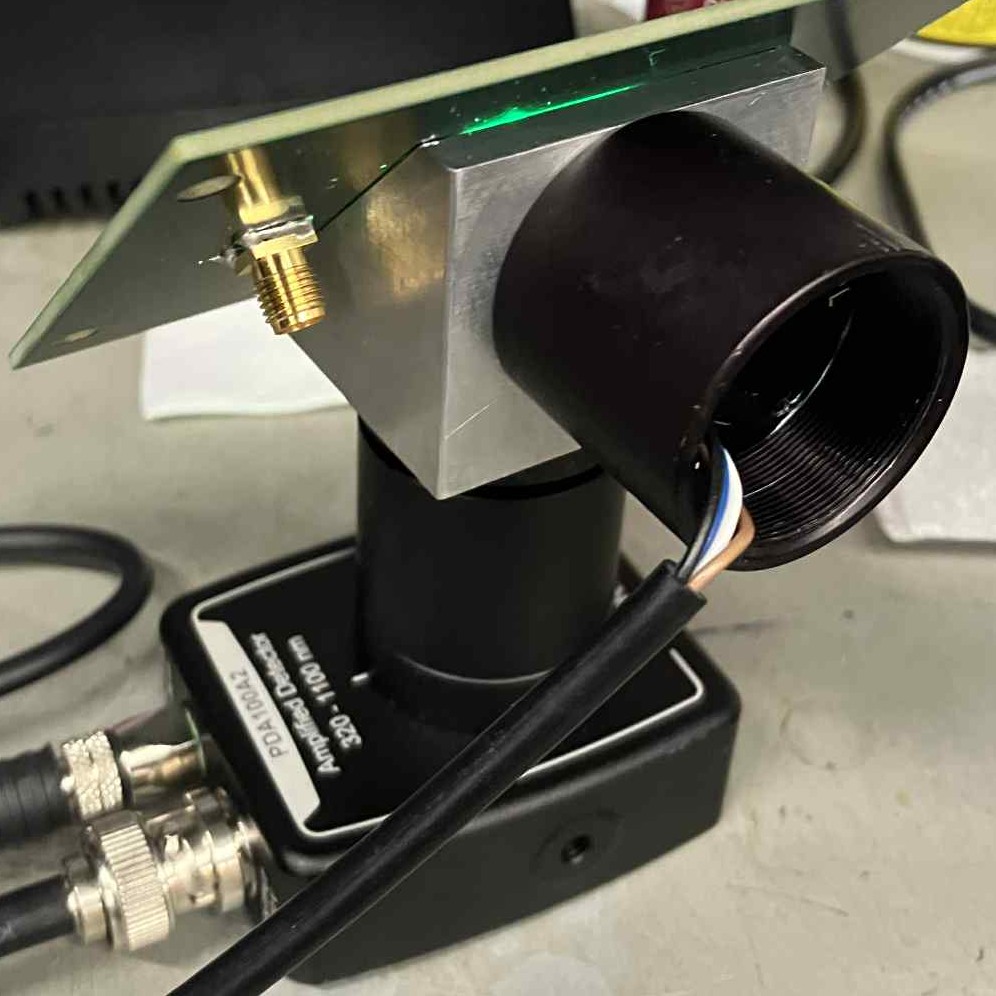}\\
    \includegraphics[width=\linewidth]{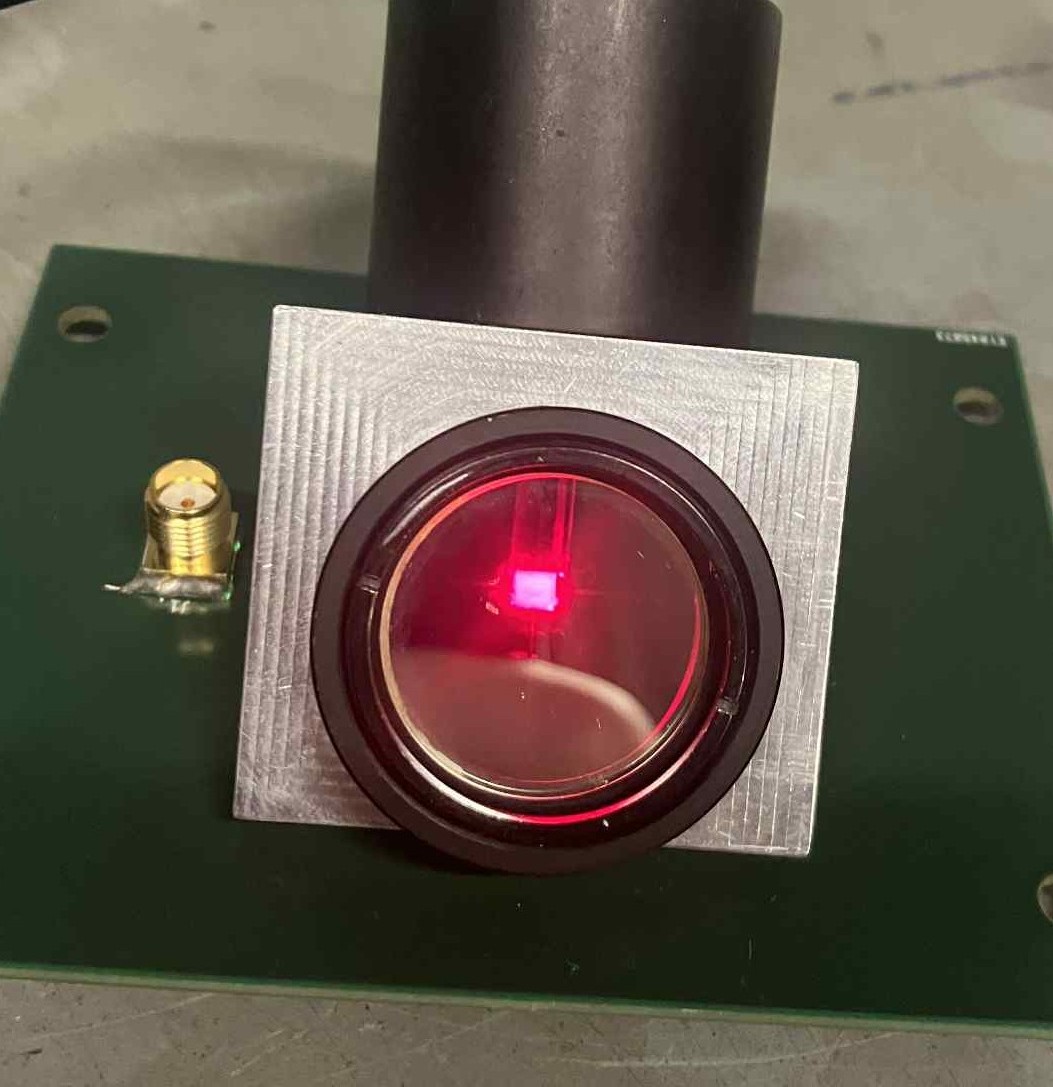}
  \end{minipage}

\caption{Photograph of the magnetometer during operation (left). Green LED excitation can be seen leaking from the device (top right), and red NV photoluminescence emitted by the diamond is readily observed through the long-pass filter (bottom right).}

  \label{fig:demo}
\end{figure}

 \begin{figure}

\includegraphics[width=100mm]{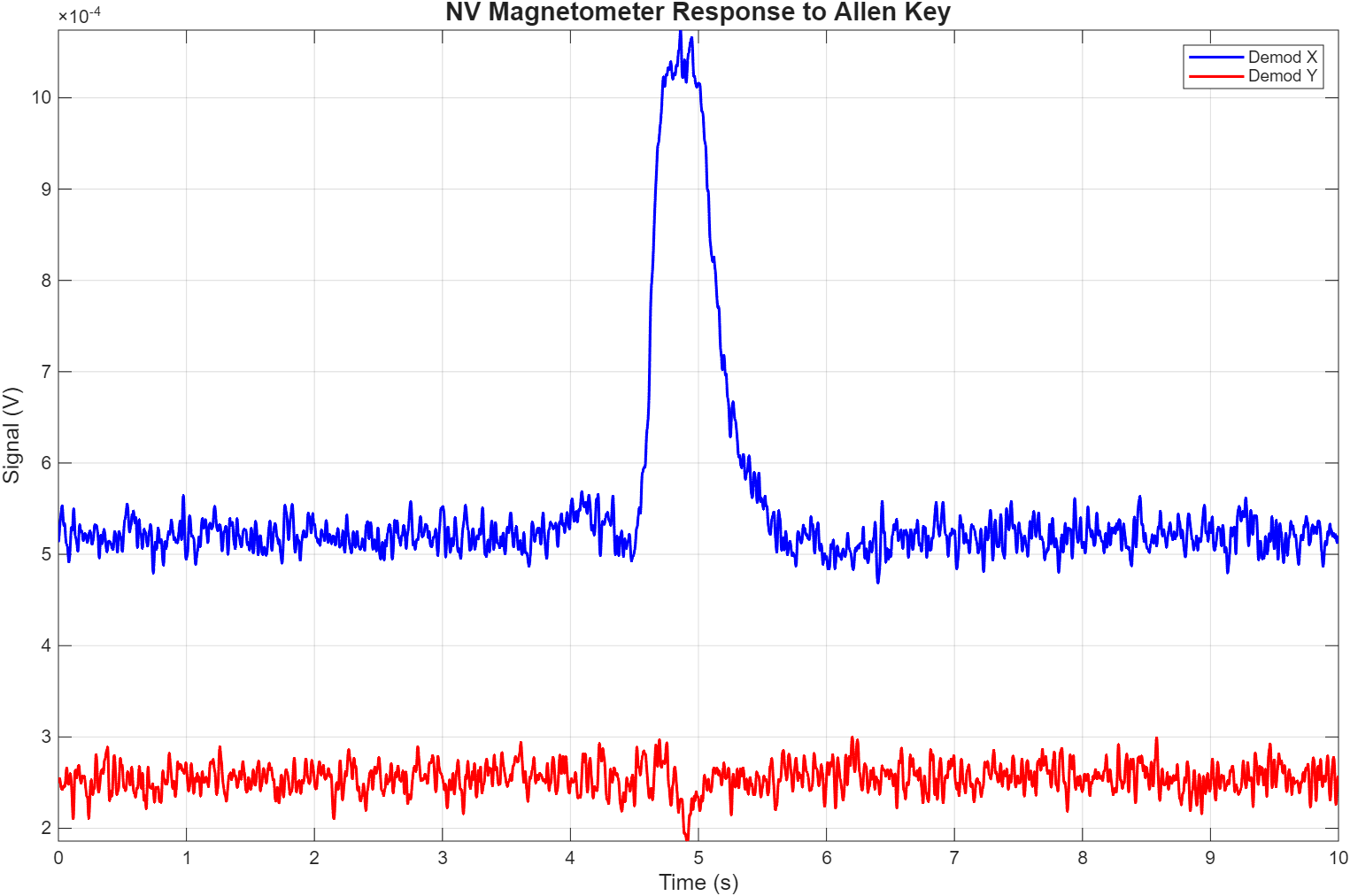}% 
\caption{\label{fig:epsart}   Time series of the demodulated PL signal when a steel Allen key is moved close to the magnetometer. The in-phase (“demod-X”) channel exhibits a spike, whereas the quadrature (“demod-Y”) channel remains near zero as expected for correctly phased lock-in detection. This measurement was performed using the LED magnetometer in its standard demonstration configuration.   }
\end{figure}

\section{ODMR Spectroscopy}

%which drives the small loop antenna patterned on the circuit board

Microwave excitation for ODMR is supplied by a compact module. In continuous-wave operation, the LED is held at constant intensity whilst the microwave frequency is swept. When the microwave field is resonant with the $\lvert m_s = 0 \rangle \!\to\! \lvert m_s = \pm 1 \rangle$ transitions, the PL decreases, producing an ODMR dip in the fluorescence. With a bias magnetic field supplied by two small permanent magnets, the degeneracy of the $\lvert m_s = \pm 1 \rangle$ states is lifted and a pair of resonances is observed.  A representative ODMR spectrum obtained with the LED magnetometer is shown in Fig. 4. The spectrum was recorded using the same operating configuration as the magnetometer demonstration (identical LED driver setting, microwave source power, and Red Pitaya lock-in configuration), ensuring that it reflects the actual working conditions of the device.  The operating parameters used to obtain Fig. 4 are shown in Table 1.

\begin{table}[h]
\centering
\caption{Operating parameters used to obtain the FM ODMR spectrum shown in Fig. 2.}
\label{tab:odmr_params}
\begin{tabular}{l c}
\hline
\hline
Parameter & Value \\
\hline
Bias field strength & $\sim 1\,\mathrm{mT}$ \\
Modulation frequency & $2.0\,\mathrm{kHz}$ \\
Modulation amplitude & $\sim 5.6\,\mathrm{MHz}$ \\
Time constant (3 dB point) & $15.29\,\mathrm{ms}$ \\
\hline
\hline
\end{tabular}
\end{table}

\begin{figure}
\includegraphics[width=100mm]{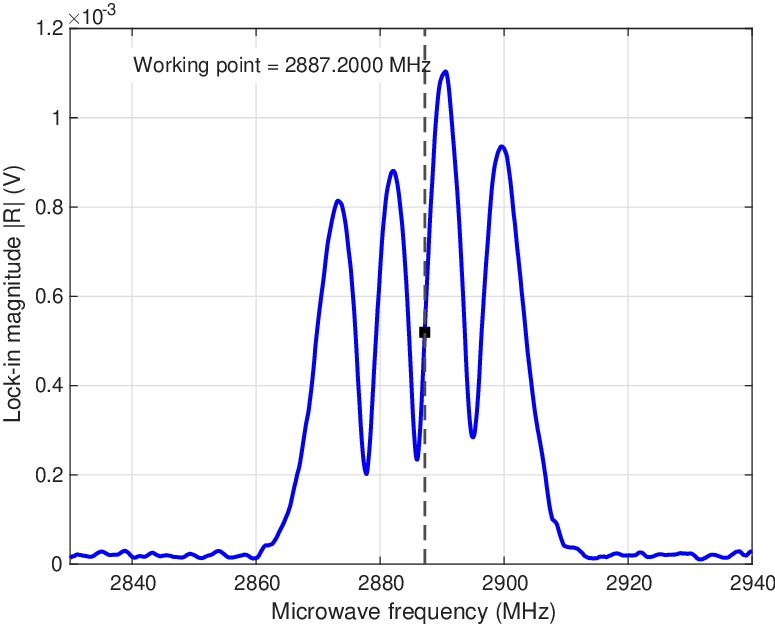}% 
\caption{\label{fig:epsart}Measured ODMR spectrum from the diamond magnetometer, acquired using frequency-modulation detection under the same operating conditions as the demonstration in Fig. 5. The observed resonances correspond to spin-state transitions between the Zeeman-split NV ground-state levels illustrated in Fig. S1.  The plotted signal corresponds to the magnitude of the lock-in response $\sqrt{X^2+Y^2}$. The dashed line marks the working point, defined as the frequency at which the ODMR amplitude exhibits its maximum slope $dV/df$.  The extrema in the FM ODMR response correspond to the steepest slopes of the underlying ODMR resonances, whose center frequencies are determined by the Zeeman splitting of the NV energy levels shown schematically in Fig. S1.}
\end{figure}

In our device, the microwave frequency is modulated and the resulting lock-in signal is therefore proportional to the derivative of the ODMR fluorescence with respect to frequency.  Although this produces derivative-like lineshapes, the resonance frequencies (and hence the Zeeman splittings) are unchanged and can be directly related to the NV energy level structure.  The lock-in amplification allows small changes in fluorescence to be measured with high signal-to-noise ratio.  For readers unfamiliar with lock-in detection, clear pedagogical treatments can be found in \cite{scofield} and in the context of NV-based ODMR measurements in Section 4D of \cite{sewani}.

\section{Magnetometry Mode and Calibration}
 
To verify quantitative operation of the magnetometer, the device was placed inside a Helmholtz coil and the coil current was varied whilst recording the demodulated lock-in response. For each current value, the magnitude of the lock-in signal  $|R| = \sqrt{X^2 + Y^2}$  was averaged over a one second interval. As shown in Fig. 5, the response varies linearly with the applied coil current over the operating range relevant for demonstrations and instructional use, confirming that the instrument provides a response which is proportional to an applied magnetic field.  The lock-in output voltage is therefore used as a proxy for magnetic field strength.

\begin{figure}
\includegraphics[width=100mm]{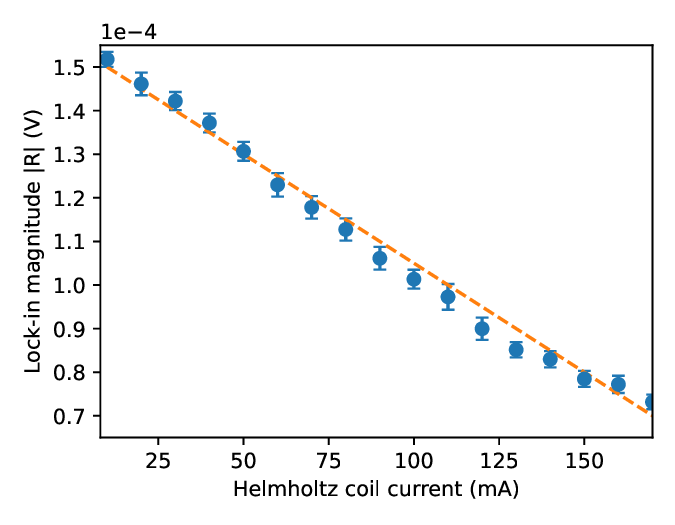}% 
\caption{\label{fig:epsart} Calibration of the magnetometer using a Helmholtz coil. The lock-in magnitude $|R| = \sqrt{X^2 + Y^2}$ is plotted as a function of coil current over the linear operating range of the device, averaged over a one-second interval.  The blue circles show the measured lock-in magnitude and the dashed orange line shows a linear fit, demonstrating that the response is proportional to an applied magnetic field.  }
\end{figure}

For more quantitative laboratory use, full ODMR spectra can be acquired by scanning the microwave frequency as described above.  Additional details on the circuit board, a complete parts list with prices, CAD files, and the sensitivity characterization of the magnetometer are provided in the Supplementary Material.

%\subsection{\label{sec:level2}Second-level heading: Formatting}

%\subsubsection{\textbf{Wide text (A level-3 head)}}
 
\section{\label{sec:level4} Conclusion}   

We have presented a portable NV center magnetometer which uses LED illumination instead of a laser, making it especially suitable for outreach and undergraduate laboratory settings. The device produces clear ODMR spectra and responds visibly to nearby magnetic objects, enabling both qualitative demonstrations and quantitative measurements.  During outreach events the magnetometer proved easy to assemble and operate, with students able to observe its response in real time. In an instructional laboratory, the system can support introductory activities, as well as more advanced exercises involving magnet alignment, bias-field estimation, or parameter optimization.

We note that although most components of the magnetometer are inexpensive, the diamond represents a significant fraction of the total cost.  Much cheaper alternatives based on ensembles of microdiamonds are possible and are widely used in other contexts. However, with the simple excitation and collection optics employed here, fluorescence would be collected from many randomly oriented crystallites, leading to significant broadening of the ODMR spectrum and reduced sensitivity.  The present design prioritizes clarity of the spectrum, robustness, and ease of interpretation over minimal cost, which is advantageous for teaching laboratories and outreach demonstrations.

%\section*{Author Declarations}

%\noindent
%The authors have no conflicts to disclose.

\begin{acknowledgments}

Alex Newman's PhD studentship was funded by an EPSRC iCASE award to UKNNL (United Kingdom National Nuclear Laboratory). This work received funding from the National Nuclear Laboratory’s Science and Technology programme (Decontamination and Decommissioning Core Science). This work is also supported by Innovate UK grant 10003146, EPSRC grant EP/V056778/1 (Prosperity Partnership with Element Six), an EPSRC Impact Acceleration Account (IAA) award, the EPSRC Q-BIOMED Hub EP/Z533191/1, the EPSRC Quantum Computation and Simulation Hub EP/T001062/1, and STFC grant number ST/W006561/1. This research is funded in part by the Gordon and Betty Moore Foundation through Grant GBMF12328, DOI 10.37807/GBMF12328. This material is based upon work supported by the Alfred P. Sloan Foundation under Grant No. G-2023-21130.

\end{acknowledgments}

%\appendix

\title{\textbf{Supplementary Material: A portable LED-based diamond magnetometer for outreach and teaching labs} 
}% 

% \affiliation{%
%Graduate School of Advanced Science and Engineering, Hiroshima University, Kagamiyama 1-3-1, Higashi Hiroshima 739-8530, Japan
%}%

\author{Hollis Williams}%
 \email{Current address: Department of Mathematics and Statistics, University of Exeter, Exeter EX4 4QF, UK}
  \altaffiliation[]{holliswilliams@hotmail.co.uk}
 \author{Alex Newman}%
\author{Stuart Graham}%
\author{Colin Stephen}%
\author{Gavin Morley}%
\email{gavin.morley@warwick.ac.uk}
\affiliation{%
 Department of Physics, University of Warwick, Coventry CV4 7AL, United Kingdom
}%

% \author{Hollis Williams}
% \altaffiliation[]{Hollis.Williams.1@warwick.ac.uk}
 
%%\author{Alex Newman}%
%\author{Stuart Graham}%
%\author{Colin Stephen}%
%\author{Gavin Morley}%
% \email{Contact author: Second.Author@institution.edu}
%\affiliation{%
% Department of Physics, University of Warwick, Coventry CV4 7AL, United Kingdom
%}%

%\affiliation{
% First affiliation for this author
%}%
%\affiliation{
% second institution for this author
%\affiliation{%



\maketitle

%\tableofcontents
 
\section*{Supplementary Material}

\section{\label{sec:level1} Background}

A nitrogen vacancy (NV) center is a lattice defect in a diamond achieved by exchanging two carbon atoms with a nitrogen atom next to a lattice vacancy \cite{nv}.  The NV center has several possible charge states, where the negatively charged state is often used as a qubit, mainly because of its spin and optical properties \cite{aslam}.  The basic effect which makes it possible to use NV centers for magnetometry is known as magnetic resonance.  The principles behind this effect are often introduced in the context of electron paramagnetic resonance (EPR) spectroscopy.  The type of resonance which is relevant for NV centers is optically detected magnetic resonance (ODMR).  As the name suggests, the main difference between EPR and ODMR is the method of read-out for the resonance.

The intrinsic magnetic moment $\boldsymbol{\mu}$ of the electron is given by 

\[ \boldsymbol{\mu} = - g_e \mu_B \hat{\textbf{S}} , \tag{1}  \]

\noindent
where $\hat{\textbf{S}}$ is the spin operator, $g_e$ is the $g$-factor for the free electron, and $\mu_B$ is the Bohr magneton.  In classical physics, the energy $U$ of a magnetic moment in an applied magnetic field $\textbf{B}$ is

\[U  = - \boldsymbol{\mu} \cdot \textbf{B} .\tag{2} \]

\noindent
This implies that if we neglect contributions from orbital angular momentum, we can promote the magnetic moment to an operator to obtain the Hamiltonian

\[   H = - g_e \mu_B   \hat{\textbf{S}} \cdot \textbf{B} .  \tag{3}\]

If the magnetic field is aligned with the $z$-direction, the energy of the magnetic moment of the electron is given by

\[   U = - g_e \mu_B B m_s ,  \tag{4}  \]

\noindent
where $B$ is the magnitude of the magnetic field and $m_s$ is the magnetic spin quantum number of the electron.  Since the total spin quantum number $S$ for an unpaired electron is $+1/2$, it follows that $m_s$ can take either the value $+1/2$ or $-1/2$, leading to two distinct energy levels, where the distance between the levels is given approximately by $g_e \mu_B B$ (shown in Fig. 1 (left)).  It can be seen that there is degeneracy when $B =0$ and that two energy levels form for $B>0$, a phenomenon known as the Zeeman effect.  If incident electromagnetic radiation (usually microwave radiation in this case) has an energy equal to the difference between the levels, absorption occurs.  In EPR, one usually keeps the microwave frequency fixed and sweeps the magnetic field strength, whereas the microwave frequency is usually swept in ODMR \cite{goldfarb, schweiger}.  The basic physics discussed above also applies to the NV center as a spin-$1$ state, although there are various complicating factors, such as the zero-field and hyperfine interactions.  The $g$-factor (or more specifically, the gyromagnetic ratio) for the electron is used in calculations for NV centers (for example, in obtaining the sensitivity result which we present in Section III).

% in the case where $S>1/2$

\begin{figure}[b]
\includegraphics[width=80mm]{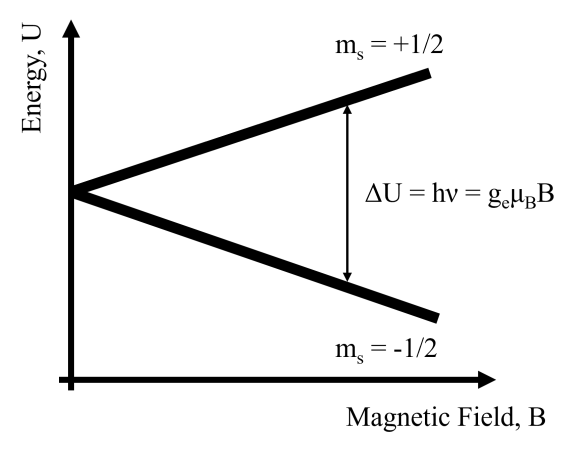}\includegraphics[width=90mm]{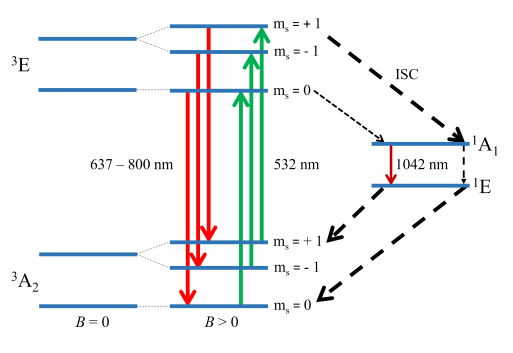}% Here is how to import EPS art
\caption{\label{fig:epsart} (Left) diagram for a spin-$1/2$ electron in an applied magnetic field $B$ showing splitting of energies between the $m_s = + 1/2$ and $m_s = - 1/2$ spin sub-levels and (right) schematic of the NVC electronic structure, with intersystem crossings and electric dipole transitions shown.}
\end{figure}

%  Reproduced from \cite{gra}

In Fig. 1 (right), we show the electronic energy level structure for the NV$^-$ center, which has electron spin $S=1$ \cite{jensen}.  The ground $^3A_2$ and excited $^3E$ energy levels shown in the figure are spin-triplet states which split into spin sub-levels with $m_s = 0$ and $m_s = \pm 1$, where the axis for the spin quantization is taken to be the symmetry axis which connects the substitutional nitrogen atom with the lattice vacancy \cite{loubser, reddy, rondin}.  There are also two spin singlets $^1A_1$ and $^1E$ \cite{rogers, acosta}.  Electron interactions lead to a zero field splitting between the spin sub-levels in the ground state by $D = 2.87$ GHz at room temperature, whereas the spin sub-levels for the first excited state are split by 1.42 GHz \cite{fuchs, neumann}.  $D$ can change depending on the temperature and varies as $dD/dT = -74$ kHz K$^{-1}$ at room temperature \cite{acosta2}.  Magnetic dipole transitions with selection rules $\Delta m_s = \pm 1$ can be stimulated using microwave radiation.  The resonances can then be read out using EPR or ODMR.

In this article, we use the standard method of ODMR with a sweep through microwave frequencies, where two resonances appear in the spectrum due to the splitting of the $m_s = \pm 1$ spin levels by the Zeeman effect \cite{taylor, bala, maze, barry}.  In the regime of sufficiently weak (but not too weak) magnetic fields, the splitting of the sub-levels is proportional to the projection of the magnetic field along the symmetry axis for the NV center \cite{rondin}.  This gives resonance frequencies in the ODMR spectrum equal to $(2.87 \pm \gamma_e B_{\text{NVC}})$ GHz, where $\gamma_e \approx 28$ GHz/T is the gyromagnetic ratio and $B_{\text{NVC}}$ is the projection of the magnetic field along the symmetry axis of the NV center.

It follows that the splitting between the two resonant peaks in the spectrum is given by $2 \gamma_e B_{\text{NVC}}$.  Since the NV center can be oriented in four ways along its symmetry axis, a total of eight resonances can often be seen when the microwave frequency is swept.  It is also possible to apply the magnetic field in certain orientations so that some of the resonances sit on top of each other.  If the field is aligned along the [1 0 0] orientation, the projection of the field along a symmetry axis is the same for all four alignments of the NV center, so that the four resonances for each side collapse into one, and only two resonant peaks are observed. 

The specific protocol which we use in our experiment is continuous-wave ODMR (cw-ODMR), where an ensemble of many NV centers in the diamond are all continuously illuminated with a green laser at the same time and driven with microwaves \cite{barry}.  This work could be extended to demonstrate pulsed ODMR.  An applied field known as the bias field with a strength between 1 and 10 mT lifts the degeneracy of the spin sub-levels and a magnetic object then causes the ODMR resonance frequencies to shift relative to the values set by the field.  A microwave frequency is fixed to a resonance frequency for the ODMR and additional fields cause a shift in the resonance frequency, which is observed as a change in the fluorescence.  Since the signal is extremely weak, one usually enhances the signal-to-noise ratio with phase-sensitive detection via a lock-in amplifier (the role of the LIA in our experiment will be played by a portable device called a Red Pitaya).

\section{\label{sec:level1}  LED Magnetometer}

\subsection{\label{sec:level1}  Design}

%RF-Consultant TPI-1005-A microwave source

In this section, we discuss the components used to build the magnetometer (these are listed in Table I along with the model, estimated cost, and suggested supplier).  The prices were checked on 29th December 2026 and there may be some variation among different vendors.  It is also possible to lower the cost by looking for used parts. The overall estimated cost of the device is around \$4500, similar to the setup described in \cite{zhang}.  In some cases, cheaper alternatives can be used to decrease the cost (for example, filtered glass could be used instead of a dichroic shortpass filter).  We also expect that some re-design of the LED part could result in a substantial price reduction.

The only two components which are not 'off the shelf' are the metal wedge-shaped cornerpiece with two threaded holes for SM1 tubes and the circuit board with a metal pad on which the diamond sits.  The former was machined in a workshop at the University of X and the latter was made by Eurocircuits.  To aid anyone that is wishing to make a similar device, we include in the Supplementary Files technical drawings and CAD files that were used to make the circuit board.  The metal pad is sized for the diamond which we use and is mainly for reflecting back fluorescence and excitation light.  Aside from this component, the setup is relatively straightforward.

 We used a $3$ mm$^2$ CVD NV diamond from Element Six which had been irradiated and annealed.  Suitable NV diamonds for the LED magnetometer could include DNV-B1 and DNV-B14 from Element Six, both of which cost around $\$2000$.  The diamond is mounted on a standard PCB pad composed of copper with a thin nickel-gold (ENIG) finish, a non-magnetic surface which does not affect the applied bias field or the ODMR measurement.  It may also be possible to use yellow diamonds if the user is able to anneal and irradiate them, where irradiation is needed to make vacancies (line widths will be broad but they should still function).  For reference, vacancies become mobile above $800^{\circ}$.  Typically, an inert atmosphere is used and the diamonds are buried in diamond grit, with annealing generally being carried out for $3$ hours at $1000^{\circ}$ C.  This temperature is high enough for good mobility but low enough that the NV complex is stable.  We provide a specific model for the SMA cables and the BNC converters used in the device, but we do not provide information on other miscellaneous screws, nuts or cables since these are readily available and the cables are recommended with the associated component by the supplier or sold with the component in question.

In constructing the device, we begin by securing the diamond to the metal pad on the circuit board with optical adhesive.  The prism is placed on top of the diamond using optical adhesive.  After this, the green LED and the longpass filter are threaded to the metal cornerpiece along with their respective SM1 tubes, where the LED is connected to the hexagonal light mixing rod.  Since emission from the diamond is not focussed, the mixing rod helps to guide the emission to the photodetector and also gives distance between the electronics and the diamond.  The cornerpiece is then placed down onto the circuit board so that the face of the light mixing rod is pressing up against the face of the prism, ready to direct light onto the diamond to reduce loss compared to an air-diamond interface.  Ideally, during this phase one should use optical adhesive to sandwich the shortpass filter between the face of the rod and the face of the prism.  However, we found the process of fitting the shortpass filter in this way to be difficult and that the filter did not have much of an effect on the sensitivity of our device, so we suggest that this step could be skipped.  

The design could also be improved so that the shortpass filter can be installed more easily.  If this stage is followed, one will also need a glass scribe or diamond scribe to score and break the filter into a piece that fits against the face of the rod, but this is cheaply available at Thorlabs or a similar supplier.  If full ''slabs'' of filter are used, more light may be directed out of the path via total internal reflection within the slab.  At this stage, one can remove the SM1 tube for the longpass filter, turn on the green LED, and observe that red light is coming from the prism via the filter.  Green light can be seen leaking out of the side of the device.  This could be part of a demonstration, as it shows that the green light from the LED and the red fluorescence can be seen directly.

The SM1 tube with the longpass filter is screwed to the amplified photodetector.  The device then naturally sits on top of the photodetector like a stand.  A neodymium block magnet is attached to the upper right and bottom left corners of the circuit board using blu-tack so that both magnets sit flat against the board.  These magnets provide the exernal bias field, which can be adjusted by changing the position and orientations of the magnets.  A laptop is connected to the Red Pitaya STEMLab using a dongle.  The Red Pitaya is a Xilinx Zynq-based combined FPGA and ARM system on chip (SoC) with capability as a portable USB oscilloscope.  Custom-made Python software enables the Red Pitaya to be used as a lock-in amplifier and users wishing to implement the digital lock-in detection and automated microwave scans can consult the scripts provided in \cite{stimpson}.  The first input port of the Red Pitaya is connected to the output of the photodetector using an SMA cable with a BNC converter.  The first output port of the Red Pitaya is connected to the FM input of the microwave source, also with an SMA cable and a BNC converter.  The custom-made circuit board is connected to the radio frequency output of the microwave source with a right-angle SMA cable.  The LED driver is connected to the other port of the photodetector and the USB port of the microwave source is connected to the Red Pitaya with a USB cable.

 \begin{table}
 \caption{\label{tab:table3} Suggested components for the LED magnetometer.}
\begin{ruledtabular}
\begin{tabular}{lcdr}
\textrm{Component}&
\textrm{Model}&
\textrm{Estimated Cost}&
\textrm{Supplier}\\
\colrule
Red Pitaya STEMLab & 125-14& \$400 &   \text{Red Pitaya}\\
Microwave source  &TPI-1005-A& \$300 & \text{RF Consultant}  \\
Dichroic 600 nm shortpass filter & 69-180& \$100 & \text{Edmund Optics} \\
Amplified photodetector & PDA100A2 & \$400 & \text{Thorlabs} \\
T-Cube LED driver & LEDD1B& \$300 & \text{Thorlabs} \\
Small neodymium cube magnet $\times 2$ & BA192& \$20 & \text{Sigel} \\
Optical adhesive & NOA 61& \$30 & \text{Norland} \\
$3$ mm$^2$ CVD NV diamond  & DNV-B1   & \$2000 & \text{Element Six} \\
SM1 tube $\times 2$ & SM1PL & \$20 & \text{Thorlabs} \\
Green LED & M530L4 & \$250 & \text{Thorlabs} \\
Cornerpiece with two SM1 threads  & CAD file & \$20 & \text{Workshop} \\
Circuit board for diamond  & CAD file & \$300 & \text{Eurocircuits} \\
Hexagonal light mixing rod  & HMR425 & \$100 & \text{Thorlabs} \\
NBK7 right angle optical prism  & 45-948 & \$70 & \text{Edmund Optics} \\
638 nm longpass filter  & DMLP638T & \$100 & \text{Thorlabs} \\
SMA cable $\times 2$  & 047-12SMPSM+ & \$20 & \text{Minicircuits} \\
Power supply for photodetector  & LDS12B & \$70 & \text{Thorlabs} \\
Right-angle SMA cable  & ACX1611-ND & \$20 & \text{Digikey} \\
SMA to BNC converter $\times 2$ & 242102 & \$10 & \text{Digikey} \\

\end{tabular}
\end{ruledtabular}
\end{table}

\subsection{\label{sec:level1}  Sensitivity}

In Fig. 2, we plot the sensitivity of the device as a function of frequency.  The main motivation for this was to see how effective the device is given its simplicity.  It is also of pedagogical interest to teach about sensitivity and parameter optimization.  As expected, the sensitivity of the device is poor in comparison with more cutting-edge NV magnetometers, with a noise floor of 1 $\mu$T/$\sqrt{\text{Hz}}$. 
 However, this value is reasonable given the simplifications made with the design of the magnetometer and the fact that a laser was not used.  Since no ODMR contrast is available when the microwaves are off-resonance or when the LED is turned off, the derived sensitivity in those cases represents the measurement noise floor rather than a physical magnetic field sensitivity.  The signal at 50 Hz is due to the main power for electrical equipment in the lab.

\begin{figure}[b]
\includegraphics[width=160mm]{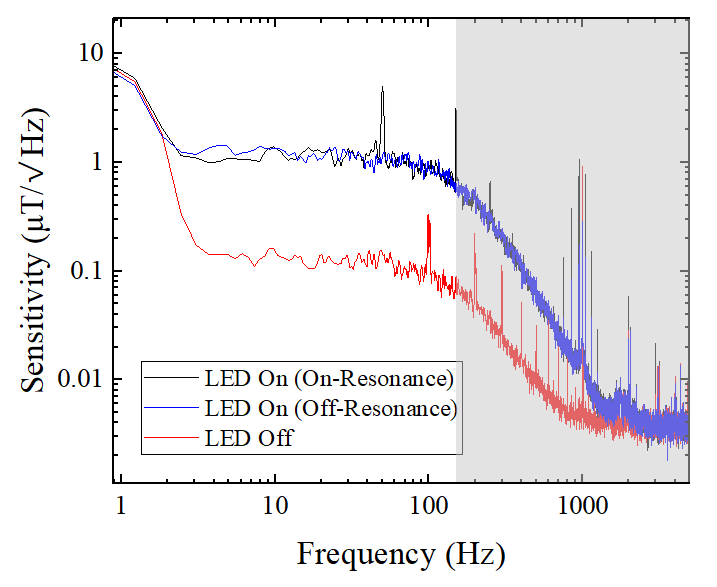}% Here is how to import EPS art
\caption{\label{fig:epsart} Plot of frequency vs sensitivity for the LED magnetometer.  The black and blue lines indicate the result when the microwave driving is done at the frequency of a resonant peak and far away from a peak, respectively.  With the LED turned off, the ODMR contrast vanishes and the sensitivity curve simply reflects the electronic noise floor of the detection chain.  The lock-in amplifier is set to reject everything above around 100 Hz using a lock-in time constant of 10 ms.  This means the equipment is not sensitive at frequencies above around 100 Hz, hence we have shaded this region of the plot gray.  }
\end{figure}

\subsection{Sample Laboratory Exercise}

This section outlines a sample laboratory exercise suitable for a
2-3 hour undergraduate teaching lab, intended to demonstrate that the
magnetometer supports hands-on experimental investigation.

\subsubsection*{Learning objectives}

After completing this guided laboratory exercise, students should be able to:
\begin{itemize}
    \item Describe the basic physical principles underlying optically detected magnetic resonance (ODMR) in NV centers.
    \item Record and qualitatively interpret ODMR spectra using lock-in detection.
    \item Estimate a magnetic field from measured resonance splittings.
    \item Explore how experimental parameters influence signal quality and measurement uncertainty.
\end{itemize}

\subsubsection*{Experimental procedure}

\begin{enumerate}
    \item With no applied bias field, students record an ODMR spectrum using the
    Red Pitaya interface and identify the zero-field splitting.
    
    \item A bias magnetic field is applied using two permanent magnets placed on
    the circuit board. Students adjust the magnet positions to resolve two
    resonance peaks corresponding to a preferred NV orientation.
    
    \item Students measure the frequency separation between the resonances and
    estimate the magnetic field strength using the relation
    $\Delta = 2 \gamma_e B_z$, accounting for the crystallographic projection
    factor appropriate for the observed NV orientation.
    
    \item The procedure is repeated for several magnet configurations to explore
    the reproducibility and uncertainty of the field estimate.
\end{enumerate}

\subsubsection*{Data analysis and discussion}

Students analyze how the resonance splitting depends on magnet placement and discuss the dominant sources of uncertainty, including linewidth, signal-to-noise
ratio, and magnetic field alignment. The role of lock-in detection in enhancing measurement sensitivity can be explored qualitatively by varying modulation
parameters within the software interface.

\subsubsection*{Extensions}

This core exercise can be extended by comparing measurements with and without frequency modulation, investigating the effect of optical or microwave power on
signal contrast, or introducing a calibrated external field source for quantitative comparison.

\subsection{\label{sec:level1}  Suggested Extensions and Project Ideas}

There are various ways that the magnetometer we developed could be studied further and used in projects for undergraduate labs.  We offer some suggestions and advice below:

\begin{itemize}
    \item Better understanding of the physics of the LED.  Relationship between the power of the LED and fluorescence.  Measurement and calibration of the LED power with regards to where the dial is positioned on the LED driver. 

   \item Parameter optimization (LED power, microwave power, modulation frequency, modulation depth).  Link between the physics of the NV center, pumping and optimization.  This is useful to understand how a lock-in amplifier works.

   \item Can the bias field be adjusted to find orientations apart from [1 0 0]?  Is it possible to obtain orientations in which all eight peaks can be seen.  This would be an easy exercise to introduce in a lab since it only involves experimenting with different positions of the block magnets on the magnetometer and inspecting the ODMR spectrum on the Red Pitaya user interface.  To achieve a [1 1 1] orientation, it may be necessary to fix the magnets at an angle so that they no longer sit flat on the circuit board.

   \item Can the design of the magnetometer be improved and made more compact?  What are the physical limitations involved in doing this and does this worsen the sensitivity?  As an example, can an LED be incorporated directly into the circuit board (this would reduce the cost of the device further) \cite{pogor}.

   \item What happens when the magnetometer is rotated in the Earth's magnetic field?
    
\end{itemize}

%\appendix

\end{document}